# Identification of Collapsed Carbon Nanotubes in High-Strength Fibres Spun from Compositionally Polydisperse Aerogels

*María Vila [a,b], Seungki Hong [c], Seung-Kyu Park [c], Anastasiia Mikhalchan [a], Bon-Cheol Ku [c], Jun Yeon Hwang [*c] and Juan J. Vilatela [*a]*

[a] IMDEA Materials Institute, c/ Eric Kandel 2, Getafe 28906, Madrid, Spain.

[b] Escuela Técnica Superior de Ingeniería de Telecomunicación (ETSIT), Universidad Rey Juan Carlos, C/Tulipán s/n, 28933 Móstoles, Spain

[c] Institute of Advanced Composite Materials, Korea Institute of Science and Technology (KIST), Jeonbuk 55324, Republic of Korea

**ABSTRACT**

Carbon Nanotubes (CNTs) of sufficiently large diameter and a few layers self-collapse into flat ribbons at atmospheric pressure, forming bundles of stacked CNTs that maximize packing and thus CNT interaction. Their improved stress transfer by shear makes collapsed CNTs ideal building blocks in macroscopic fibers of CNTs with high-performance longitudinal properties, particularly high tensile properties as reinforcing fibres. This work introduces cross-sectional transmission electron microscopy of FIB-milled samples as a way to univocally identify collapsed

CNTs and to determine the full population of different CNTs in macroscopic fibers produced by spinning from floating catalyst chemical vapour deposition. We show that close proximity in bundles is a major driver for collapse and that CNT "stoutness" (number of layers/diameter), which dominates the collapse onset, is controlled by the growth promotor. Despite differences in decomposition route, different C precursors lead to similar distributions of the ratio layers/diameter. The synthesis conditions in this study give a maximum fraction of collapsed CNTs of 70% when using selenium as promotor, corresponding to an average of 0.25 layer/nm.

**INTRODUCTION**

Like a macroscopic tube, carbon nanotubes (CNTs) of large diameter and small thickness (i.e. few layers) can have sufficiently low flexural rigidity to collapse radially inwards into flat, ribbon-like structures, and remain stable due to van der Waals interactions between opposite sides of the inner CNT layer in the collapsed state[1]. The collapse of CNTs has been widely studied by molecular-dynamic simulations and experiments, including studies on changes in their theoretical electronic properties[1–4] and recent determination of their Raman spectrum.[5] Continuum elasticity theory can be used to determine buckling pressure for the collapse of individual CNTs of different morphology.[6] When in bundles, CNT collapse can be even more energetically favorable as a consequence of better packing of flat CNT compared to round CNTs, with the pressure exerted by surrounding CNTs in a bundle analogous to a capillary force.[7,8]

Collapsed CNTs hold a special place in the quest to produce continuous macroscopic CNT fibers with high-performance mechanical properties.[9,10] As fibrillary solids, the strength of CNT fibers is proportional to the total area of contact between CNTs (assuming perfect alignment) and depends on the stress transfer in shear between neighbouring nanotubes.[11] Collapsed CNTs form bundles where the planar CNTs form stacks, reaching high packing fractions of around 0.95. In comparison, non-collapsed CNTs hexagonally-arranged lead to packing efficiencies below 0.75 even after polygonisation.[9] Furthermore, simulation work suggests that the friction coefficient between CNTs increases upon collapse because of changes in the corrugation potential of the graphitic layer.[12–14] The synthesis of long, large-diameter CNTs capable of forming stacks while in collapsed state has been considered key to achieve high mechanical properties in macroscopic CNT fibres since early work on FCCVD synthesis.[11] Indeed, various reports show that fibers with some evidence of collapsed CNTs can reach tensile strength of 2GPa/SG,[11,15,16] above some commercial carbon fibers, or thermal conductivity above that of copper.[17] More recently, fibers with collapsed CNTs were shown to stabilize intercalants, enabling large increases in electrical conductivity through hole doping via the formation of air-stable intercalation compounds.[18]

Despite the interest in producing macroscopic ensembles with a high fraction of collapsed CNTs, the challenge has proven elusive. One reason is the difficulty to distinguish the subtle molecular-scale differences between collapsed and non-collapsed CNTs, particularly using methods that can be applied to macroscopic fibers with millions of CNTs. Large areas can be probed with Raman spectroscopy, for example, but it provides inconclusive evidence of CNT collapse: the general lack of radial breathing modes (RBM) can be due to multiple reasons other than collapse, and even the observation of resolved transverse and longitudinal optical tangential modes ($G^-$, $G^+$) in the absence of RBS may be due to round double-walled CNTs. This difficulty is compounded by the fact that the process used to synthesize fibers with evidence of collapsed CNTs, floating catalyst chemical vapor deposition (FCCVD),[19] produces inherently very polydisperse samples in terms of the CNT diameter and number of layers[20] as a consequence of a wide catalyst particle size distribution[21]. Such polydispersity means the spun aerogels are not formed *exclusively* of single- or multi-layer CNTs and are instead comprised of a population of CNTs with different diameter, number of layers, and different cross-sectional morphology (e.g. round, polygonised, collapsed).

A promising strategy could be TEM analysis of the thinnest lamellas (slices) cut normal to the fibre's axis that resolves the transversal cross-sections of CNTs and their bundles at the highest resolution. Transversal TEM technique has been mainly used for the visual representation of high-densified cross-sections with reduced inter-bundle distances in CNT fibres subjected to stretching post-treatments.[22,23] Cho *et al.* noted that CNTs became more elliptic (through enhanced eccentricity from 0.46 to 0.83) with reduced intra-bundle voids in stretched CNT fibres.[24] Similarly, Jolowsky *et al.* observed DWCNTs with diameter > 5 nm in either circular or collapsed state in the composite with 80%-stretched CNT sheet, with notable increase in the packing of stacked CNT with stretching ratio.[25] In a recent example, a transversal TEM image has been shown for FCCVD-spun CNT fibres to confirm the presence of few-walled CNTs[26].

In this work, we use the transversal TEM method to observe cross-sections of CNTs in bundles of macroscopic fibers, enabling not only univocal identification of collapsed CNTs in bundles, but also accurate quantitative determination of the distribution of CNT diameters and number of layers. Equipped with this tool, we demonstrate the importance of bundling and low CNT "stoutness" (number of layers/diameter) to induce self-collapse at atmospheric pressure. Finally, the paper shows that CNT populations can be shifted through changes in the FCCVD synthesis conditions, with the fraction of collapsed CNTs increasing through the choice of promotor but being less sensitive to carbon source.

**Experimental**

**Synthesis**

CNT aerogel filaments were synthesized in the gas phase by floating catalyst Chemical Vapor Deposition (FCCVD)[19]. As schematically represented in Figure 1a, millimeter long CNTs are continuously drawn from the furnace forming 10 micron-diameter aerogel filaments that are spun directly onto a spool. The study is centered on a type of CNT fiber of predominantly collapsed CNTs, exhibiting high longitudinal tensile strength (>2 GPa/SG), modulus (>100 GPa/SG), and electrical conductivity (3 x $10^5$ S/m)[15]. An example of a CNT fiber is included in Figure 1b. These samples were synthesized at 1300 $^0$C in a hydrogen atmosphere, using toluene as a carbon source (C), thiophene as a sulfur (S) promoter and ferrocene (Fe) as a catalyst source, at atomic ratios (S/C = 3.3 x $10^{-3}$, Fe/C = 9 x $10^{-4}$). Samples were spun at a rate of 28 m/min. For comparison, samples produced using a different carbon source (butanol) or a different promotor (Se)[27] were are also produced. They had the same nominal S/C ratio and Fe/C ratios. In this work, samples for conventional TEM analysis were prepared by placing the individual CNT aerogel filaments without densification (low-density "open" aerogel)[20] on TEM grid. Those analyzed by transversal TEM were densified with a volatile solvent to form fibres, as reported elsewhere.[15,28]

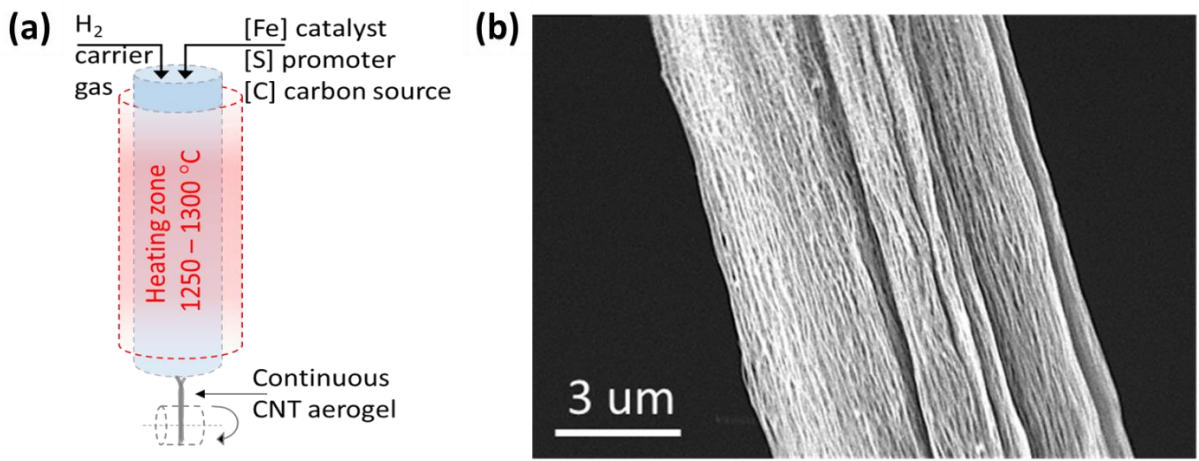

**Figure 1.** CNT fibre produced by FCCVD. a) Simplified schematic of the FCCVD synthesis process. b) CNT aerogel densified into a continuous fibre of aligned and aggregated CNT bundles.

**CNT Filament Characterization and Sample Preparation**

The samples for transversal TEM analysis were prepared by slicing a densified CNT fibre using a Focused Ion Beam (FIB) microscope (Thermo Fisher Scientific, SCIOS Dual-beam system). The fibre cross-section milling was done after Pt deposition on the CNT fibre material (approximately 2 μm) to protect the specific area of interest from ion beam damage. A cross-section slice (lamella) of CNT fibre (around 10 μm width) was prepared by means of progressive focused ion beam milling and transferred to an Omniprobe Lift-out Copper grid. Ion beam milling was first performed with an accelerating voltage of 30kV and beam current range 5nA to 1nA. After lifting out the specific area of interest and attaching to the TEM grid, ion beam milling at 30kV with 0.3nA was done to reduce sample thickness to less than 200nm. Finally, further sample cleaning was done at 5kV with 48pA until reaching 100nm sample thickness. Examples of electron micrographs during the preparation process are included in Figure S1, and the FIB ion beam milling conditions are summarized in Table 1.

To discard sample damage or modification due to FIB cross – sectioning, a subset of CNT fibers were embedded in a polymer matrix before ion beam milling. They showed the same structural features as those without polymer. Figure S2 shows examples of samples used for transversal TEM observation.

Note that the relatively high degree of alignment of the CNTs (see SAXS azimuthal profile in Figure S3) ensures that the analysis of well-defined cross-sections of CNT bundles provides accurate determination of CNT diameter. High resolution TEM images were obtained by Thermo Fisher Scientific-Titan G3 with Cs-corrector and monochromator at 80 kV. For precise cross-sectional analysis, the tilting experiment was explored.

Table 1. FIB ion beam milling parameters for preparation of cross-section lamella for transversal TEM analysis

| Progressive thinning, nA / kV | Final thinning, nA/kV | Final cleaning, pA / kV |
|---|---|---|
| 5 - 1 / 30 | 0.3 / 30 | 48 / 5 |

CNT alignment was determined from 2D small-angle X-ray scattering (SAXS) patterns from individual filaments collected at the NCD-SWEET beamline of Alba Synchrotron lightsource with radiation wavelength of 1Å. Sample-to-detector distance and other parameters were calibrated using reference materials. Data were processed as detailed in previous work[29].

**Results and Discussion**

**Conventional TEM imaging**

The materials in this study consist of continuous, macroscopic fibers of CNTs produced as aerogels in a vertical FCCVD reactor. Collecting the material in the aerogel form and depositing it directly onto TEM grids preserves the open network structure of CNTs in the material, enabling direct observation of low-density areas of the sample by conventional TEM.[20] Collapsed CNTs can be identified by their planar cross section. Figure 1a shows an example of a large number of collapsed CNTs in the sample, with obvious collapse features. Rapid identification of CNTs as collapsed is usually done when their buckled structure is exposed either through a slight twist angle or where they join a bundle. In the example in Figure 2a, collapsed CNTs are particularly easy to identify through several points where they fold onto themselves.[27] Another example is shown in Figure S4. However, these clear collapse features are generally hard to observe. A CNT may only be collapsed in some areas, thus requiring additional observation at different tilt angles to discriminate different morphologies. Most CNTs in the material are bundled in aggregates of CNTs with different size and number of layers. In fact, significant efforts are devoted to increasing their packing as a way to maximize axial fiber properties. As shown in the example in Figure 2b, it is difficult to individually resolve the CNTs in typical bundles. Sections of individual CNTs are often exposed in CNTs splayed from bundles, but determining even their basic molecular features requires very time-consuming high-resolution imaging. Figure 2c presents an example of a TEM image of a suspected collapsed CNT. It is very wide, with diameter beyond reported thresholds for collapse of 2-3 walled carbon nanotubes[30], but because the observed "diameter" is a 2D projection, this imaging mode does not by itself confirm that the CNT is collapsed. Using high-resolution imaging it can be established that it is collapsed based on the FFT of the image, which shows two sets of six – fold graphitic patterns. Each spot in the FFT image is split in two,[31] related to the two opposing graphitic walls of the tube and indicative of a chiral tube in which all walls have similar chiral angle. For comparison, non-collapsed CNTs found in the material produce a FFT with Fraunhofer streaking characteristic of the high curvature of a non-collapsed tubes[32,33] as their features are extended in two dimensions[34] (Figure 2d). These differences are also directly visible in the Moiré patterns arising from superposition of the hexagonal network of the different graphene layers observed under high resolution.[35]

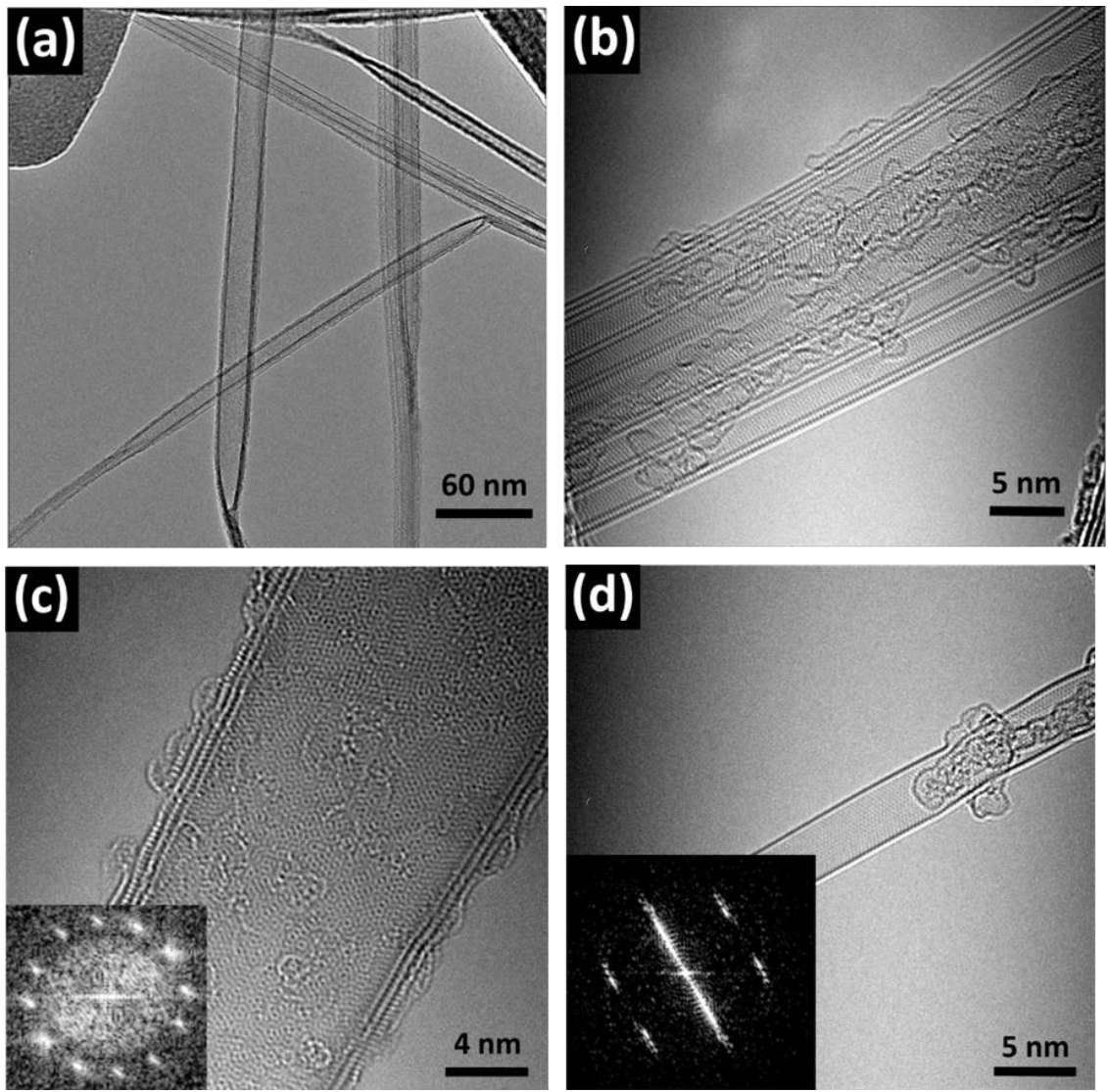

**Figure 2.** Conventional TEM images of different CNTs. (a) CNT bundle showing several collapsed tubes and (b) small - diameter CNTs in a bundle. (c) High - resolution TEM image of an isolated collapsed CNT and its corresponding FFT image as inset, (d) Non - collapsed tube with its corresponding FFT.

**Cross – sectional TEM images**

An alternative method to analyze CNTs in macroscopic fibers is to exploit the mechanical robustness of the network structure to produce transversal sections of CNT aerogels by FIB milling. Because the CNTs are predominantly aligned along the fiber axis (according to SAXS, Figure S3, Supplementary Material), transversal slices of the fiber directly expose the cross section most of the CNTs, as well as other details of the bundle structure. Figure 3 presents examples of TEM micrographs. This mode of observation of CNT fiber aerogels provides new insight into their molecular structure in terms of CNT eccentricity, fraction of collapsed CNTs, packing efficiency, and in general provides the first means for accurate determination of the distribution of CNT diameters and number of layers in these systems. These parameters were determined from analyses of several CNT fiber cross sections, from which 93 CNTs could be clearly resolved.

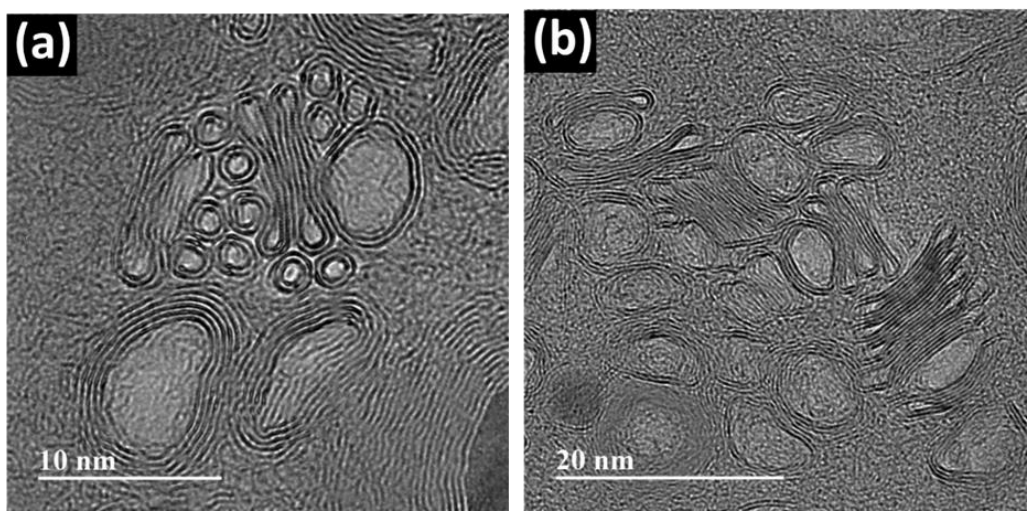

**Figure 3.** Transversal TEM images.

A first aspect that stands out from cross–sectional TEM images of CNT bundles is not only the wide range of CNT diameters, but the wide spectrum of CNT morphologies, from fully collapsed to elliptical to completely round. These features would be imperceptible through conventional TEM observation.

We characterize these morphological features through the nanotube eccentricity:

$$e = \frac{\sqrt{a^2-b^2}}{a} \qquad (1)$$

where 2a and 2b are the large and short axis respectively. Note that *e = 0* for circular sections and *e = 1* for fully collapsed tubes. Figure 4a presents the distribution of eccentricity for the CNTs analyzed, now with the confidence that collapsed CNTs can be readily identified. The distribution shows indeed the predominance of collapsed tubes (40%) over circular ones (18%). There is in addition a large fraction of elliptical CNTs (42%), and the majority have high eccentricity, with 40% of those having *e > 0.5*, for example. They are likely in a metastable state, susceptible to collapse under external pressure.

Next, we determined the full distribution of the inner and outer diameters of the CNTs, now accurately determined for elliptical and collapsed tubes from their directly-observed perimeters.

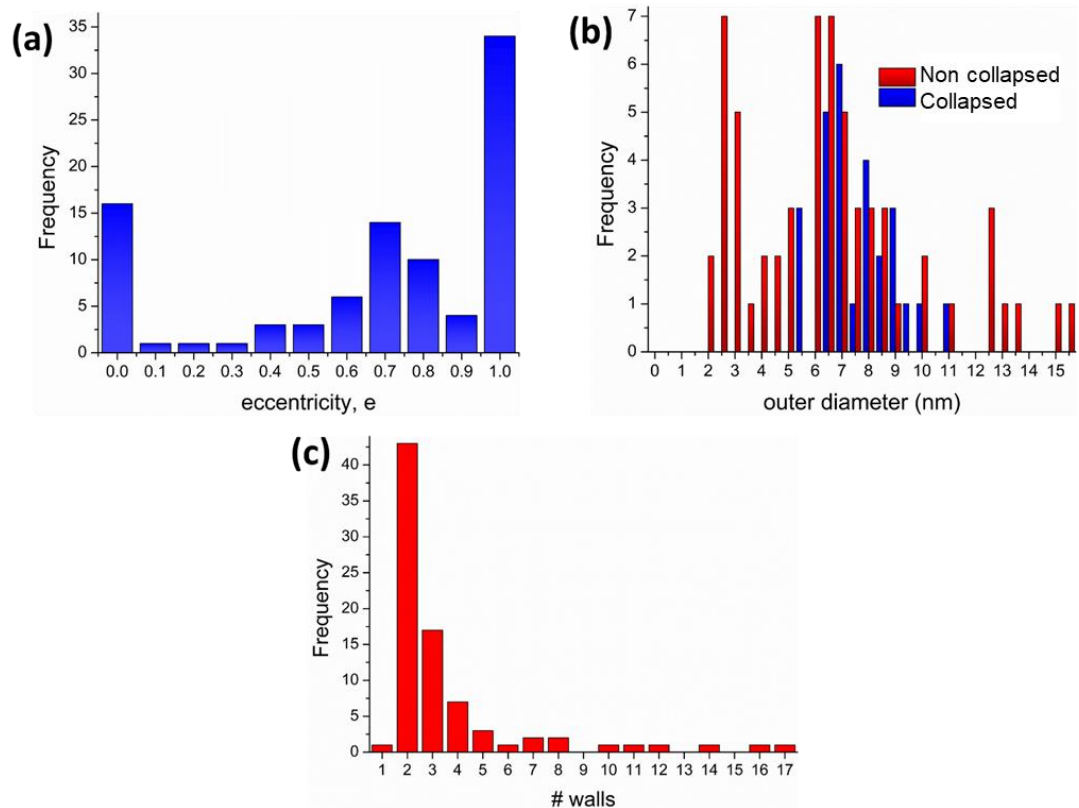

**Figure 4.** CNT morphology determined from TEM observation of cross-sections. Distributions of (a) eccentricity, (b) outer diameter calculated from observed and (c) number of walls (layers).

Both distributions (Figures 4a and S5) are broad, in the ranges 1- 9 nm and 2 – 15nm for the inner and outer diameters, respectively. This broad range reflects more than the distribution of the number of CNT layers, which is in fact much narrower, with 52.5% of DWNTs, and 75% having less than 5 layers (Figure 4c and S6,S7).

Marked in the diameter distribution are also CNTs that are collapsed. As expected, they correspond to those with a large outer diameter and a small number of layers – therefore having a larger inner diameter-. For collapsed CNTs, average outer diameter of 7.6 nm and inner diameter of 6.6 nm were obtained.

To rationalize the different CNT morphologies observed and identify descriptors of the combined effect of diameter and number of layers on CNT eccentricity, it is instructive to consider the mechanics of CNT collapse.

**Flexural Rigidity and buckling pressure**

From continuum mechanics, the effective bending stiffness (or flexural rigidity) of a single walled tube, i.e. the resistance to changes in the curvature of a single shell[4], is given by:

$$D_0 = \frac{Yt^3}{12(1-v^2)} \quad (2)$$

where $Y$ is the effective Young´s modulus, $v$ is Poisson´s ratio and $t$ is the shell thickness. This expression can be applied to single-walled nanotubes with appropriate selection of the shell thickness. Instead of the interlayer spacing of 0.34nm, the effective wall thickness of the SWNT is related to the atomic-scale wall resistance to in-plane bond stretching and in-plane angle bending, with a value between 0.066 and 0.089 nm.[2,4,36] With a reduced thickness, the corresponding modulus $Y$ = 5500 GPa, whereas the Poisson ratio remains unaltered at $v$ = 0.19.[2] Although not without controversy, this model can be expanded to MWCNTs by taking the effective thickness of $h = Nt$, where $N$ is the number of layers.[3] It then follows that CNT bending stiffness scales as $D \propto N^3$. Figure 5a shows the calculated values of flexural rigidity against number of layers for the CNT types found from TEM observation (assuming there are round). The values span from 1eV to 3000 eV, and highlight the fact that a relatively narrow distribution of the number of layers and diameters (Figure 5b) can propagate into an enormous difference in the bending stiffness of the constituent molecules of the fibers.

With the view of describing CNT packing efficiency in bundles, a more adequate descriptor than $D_0$ can be extracted from analysis of the mechanics of collapse. From continuum elasticity, the collapse buckling pressure for an isolated CNT collapse is given by[6]:

$$P_c = \frac{2Y}{1-v^2}\left(\frac{Nt}{d}\right)^3 = 24D_0\left(\frac{N}{d}\right)^3 \tag{3}$$

This equation predicts the pressure necessary for the collapse of CNTs based on the rigidity of the graphitic shell ($D_0$) and the shape of the CNT in terms of $N$ and its outer diameter ($d$). It contains the ratio of $N/d$ as the key descriptor for the collapse of the CNTs, capturing the notion that tubes with larger diameter and more layers collapse more easily.

In Figure 5c we plot values of $N/d$ for the CNTs observed by transversal TEM. The distribution shows a peak at low values of $N/d$, corresponding to collapsed CNTs, and higher values for non-collapsed tubes. The threshold for collapsed CNTs obtained from this data is approximately $N/d = 0.5$ nm$^{-1}$. This is much higher than the value corresponding to self-collapse at atmospheric pressure $N/d_{atm} = 0.027$ nm$^{-1}$, the experimental data implying that collapse occurs at much lower pressure than predicted by equation (3). Such discrepancy is due to the fact that the data corresponds to CNTs that are not isolated, but surrounded by other CNTs in bundles, which strongly favors the collapse process.[6] Pugno et al.[7,8] proposed that the presence of neighbouring nanotubes in a bundle is analogous to the action of a liquid tension, exerted through the surface energy of the nanotubes (γ). The collapse pressure for a CNT in a bundle can then be expressed as:

$$P_c = 24 D_0 \frac{N^\alpha}{d^3} - \frac{2\gamma}{d} \qquad (4)$$

where $1 \leq \alpha \leq 3$, and $\alpha = 3$ for perfect bonding between walls, and $\alpha = 1$ for independent walls. For an intermediate coupling between walls, the reported value of $N/d \sim 0.4$ nm$^{-1}$ (threshold diameter for double-wall CNTs ~ 5.4 nm) is remarkably close to the value of 0.5 nm$^{-1}$ obtained here (Figure 5c). The experimental data are also in agreement with recent atomistic simulations of monodisperse bundles showing a critical $N/d$ ratio in the range 0.29 – 0.5 nm$^{-1}$.[37] Note also that an implication from equation (4) is that atmospheric pressure has a minor effect on collapse compared to geometric factors of the constituent CNTs and their arrangement in bundles.

**Implications for CNT growth through floating catalyst CVD**

A point of interest stemming from the direct observation of CNT cross sections by TEM is the width of the distributions of diameter and number of layers, i.e., the morphological polydispersity of the 1D constituents. The shape of the diameter distribution is comparable to that other inorganic 1D nanostructures produced by FCCVD,[38] but very wide for the goal of controlling electronic properties of constituent CNTs or to improve packing of CNTs in bundles. This is partly due to the fact that the synthesis process includes the formation of catalyst particles from a precursor, thus itself producing a wide distribution of particles sizes.[21] Nevertheless, the concentration of promotor in the reaction also plays a significant role.

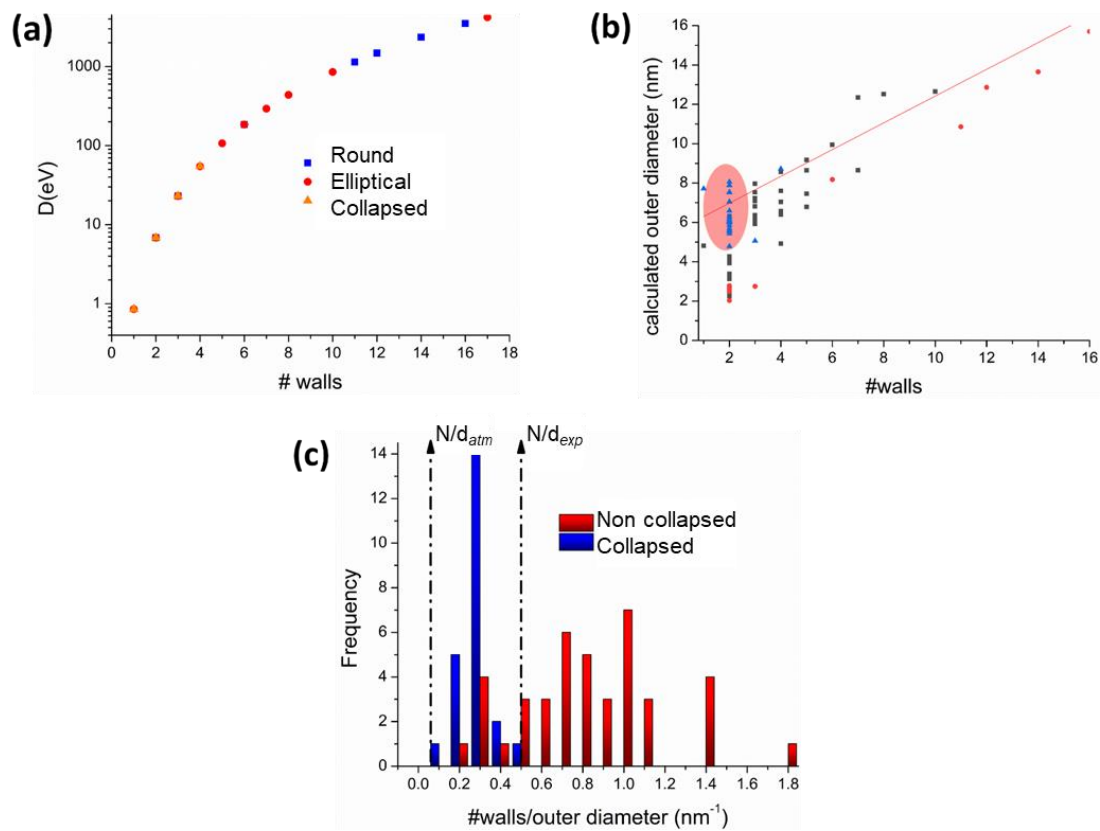

**Figure 5.** Analysis of collapsed CNTs from TEM observation of cross-sections. (a) Distribution of bending stiffness. (b) Equivalent outer diameter and number of layers determined experimentally and linear relation: $d_{outer} = d_{inner} + 0.68(n-1)$. (c) Ratio of number of walls/diameter.

Figure 6 presents a comparison of histograms of *N/d* for CNT fiber samples produced with two different carbon sources: toluene (Figure 6a) or butanol (Figure 6b) and two different promotors (S or Se), but with an approximately constant ratio of catalyst precursor to C. Examples of micrographs are included in Figure S8. The data show that the choice between these C sources has minor effect on the distributions of *N/D*. This observation contrasts with the significant differences in the bulk properties of the two fibers: those made from toluene are around twice as strong and 100 times more conducting for similar degrees of alignment, for example.[15] Moreover, we note that the two *N/d* distributions are similar despite expected differences in the decomposition route and temperature of the C sources, suggesting that adjusting decomposition temperatures routes to catalyst availability may not be a suitable strategy for molecular control. Instead, these observations support the view that the size distribution of *active* catalyst particles is largely controlled by the atomic ratio of precursor/growth promotor[39], as observed previously by analysis of catalyst particle formation[21]. Indeed, using a different promotor, for example Se, produces a significant difference in the *N/d* distribution. As shown in Figure 6c, the distribution for Se is narrower and centered at lower values, in other words, Se leads to few-walled larger diameter CNTs.

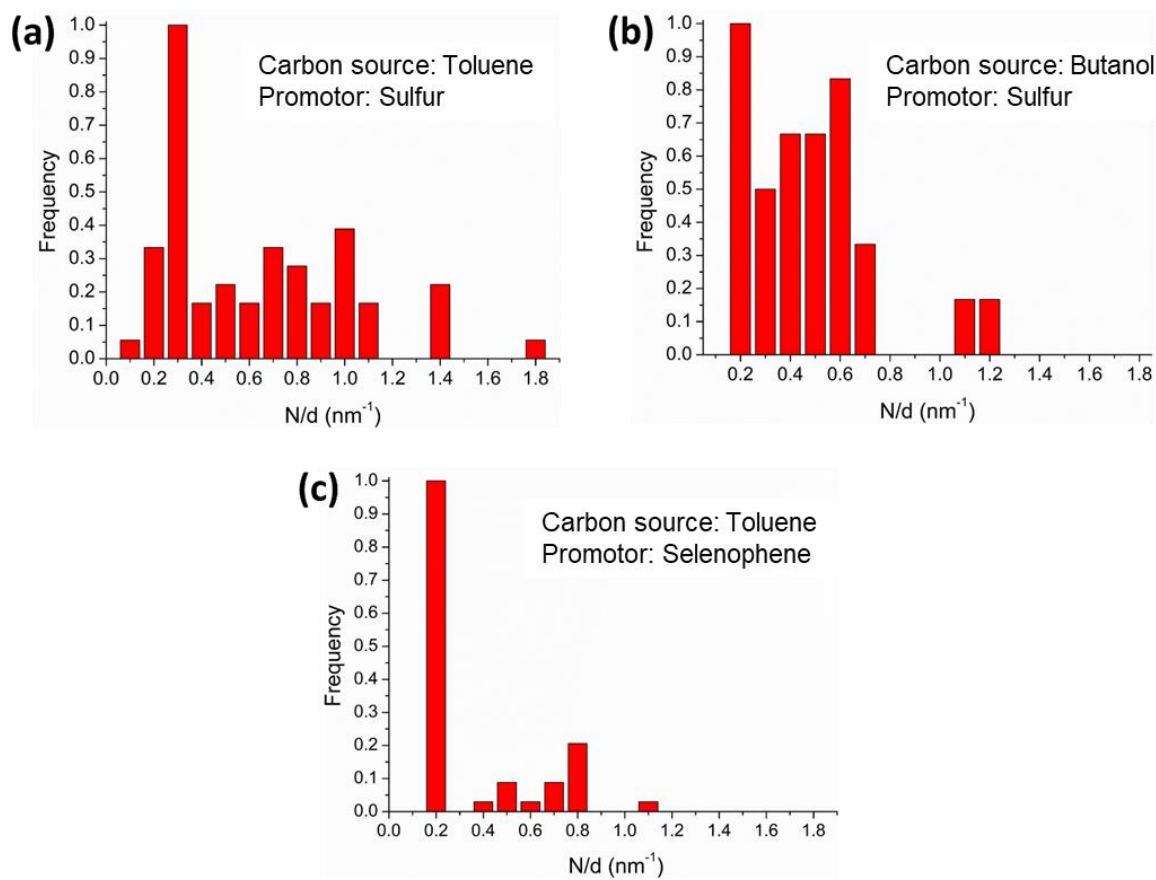

**Figure 6.** Histograms of N/d for CNT fibers produced from different carbon sources (toluene and butanol) and different growth promotor (Se and S).

In a previous work, it was proposed that *N/d* is related to the solubility of the promotor in active catalyst particles[26]. More precisely, equating the rate of incoming and outgoing carbon CNT growth suggest that the parameter *N/d* is proportional to the carbon concentration in the growth front of the ternary alloy of the nascent CNTs[27]. These new results confirm the large differences observed for CNT growth in the Fe-S-C or Fe-Se-C systems. However, the new insight gained from transversal TEM also points to the active catalyst particles having a range of Fe-S-C contents, as inferred from the fact that the CNTs have a marked *N/d* distribution instead of being single-valued (see line Figure 5b). This implies that further control over CNT morphology in CNT fibers, for instance to favor collapsed CNTs, requires not only selected size[40] but also control of the chemical composition of the aerosol of catalyst particles.

**Conclusions**

This work introduces a method for analysis of molecular composition of fibers of CNTs and other macro-ensembles of 1D nanostructures, based on TEM imaging of FIB-milled cross sections of bundles. This mode of observation of CNT fiber aerogels provides quantitative information on their molecular structure in terms of CNT eccentricity, fraction of collapsed CNTs, packing efficiency, and in general provides the first means for accurate determination of the distribution of CNT diameters and number of layers in these systems. The results show a wide distribution of diameters and number of layers, leading to the coexistence of collapsed, predominantly round or elliptical CNTs in the same bundles. These morphologies are governed by the CNT stoutness, that is, the ratio of number of layers/diameter. Comparison of materials produced under different synthesis conditions shows that the distribution of N/D is mainly controlled by the choice of the growth promotor and its effect on the distribution of active catalyst particles. Materials produced with equal ratios of promotor/carbon but from different precursors have similar N/D distributions.

From analysis of the maximum N/D values for collapse, the results also show that CNT collapse largely occurs through bundling. This suggest that methods to control aerogel formation and assembly into fibers could affect the final state of the CNTs, and could now be analysed with the method introduced here.

## ASSOCIATED CONTENT

The Supporting Information is available free of charge at …

**Supporting Information**: FIB cross-sectioning process, examples of TEM images, 2D SAXS pattern, determination of equivalent diameter and number of layers (PDF).

## AUTHOR INFORMATION

**Corresponding Author**

* Juan José Vilatela

IMDEA Materials Institute, C/ Eric Kandel 2, Getafe 28906, Madrid, Spain

juanjose.vilatela@imdea.org

* Junyeon Hwang

Institute of Advanced Composite Materials, Korea Institute of Science and Technology (KIST), Jeonbuk 55324, Republic of Korea

Junyeon.hwang@kist.re.kr

**Author Contributions**

The manuscript was written through contributions of all authors. All authors have given approval to the final version of the manuscript.

## ACKNOWLEDGMENT

The authors are grateful for generous financial support provided by the European Union Horizon 2020 Program under grant agreements 678565 (ERC-STEM), 797176 (ENERYARN), by MINECO (RyC-2014-15115, HYNANOSC RTI2018-099504-A-C22), the Madrid Regional Government (program "Atracción de Talento Investigador", 2017-T2/IND-5568 and FOTOART-CM P2018/NMT-4367), by the Air Force Office of Scientific Research of the United States (NANOYARN FA9550-18-1-7016) and by the Carbon Hub. J.Y.H is partially supported by the Korea Institute of Science and Technology (KIST) open research program (2E31332). The SAXS-WAXS experiments were performed at NCD-SWEET beam line at ALBA synchrotron with the collaboration of ALBA staff.

**Notes and references**